\documentclass[prb,aps,showpacs,unsortedaddress,superscriptaddress,amsmath,amssymb,floatfix,twocolumn]{revtex4-1}
\usepackage{epsfig}
\usepackage{subfig}
\usepackage{graphicx}
\usepackage{amssymb}
\usepackage{amsmath}
\usepackage{amsthm}
\usepackage{wasysym}
\usepackage{hyperref}
\usepackage[usenames]{color}
\begin{document}

\title{Role of work function distribution on field emission effects}
\author{Nandan Pakhira}
\affiliation{Department of Physics, Kazi Nazrul University, Asansol, West Bengal 713340, India}
\author{Rajib Mahato}
\affiliation{Department of Physics, Kazi Nazrul University, Asansol, West Bengal 713340, India}
\affiliation{Central Electronics Engineering Research Institute, Pilani, Rajasthan 333031, India}

\begin{abstract}
Field emission effect is the emission of electrons from a cold metallic surface in the presence of an electric field. The emission current exponentially depends on the work function 
of the metallic surface. In this work we consider the role of work function distribution on the field emission current. The work function distribution can arise due to 
nano-scale inhomogeneities of the surface as well as for collection of nano-particles with size distribution. We consider both Gaussian distribution as well as log-normal distribution. 
For Gaussian distribution, the field emission current, $J_{\textrm{av}}$, averaged over work function distribution shows Gaussian dependence, $J_{\textrm{av}}\propto \exp(\alpha\sigma^{2})$,
where $\sigma$ is the width of the work function distribution and $\alpha$ is a fitting parameter. For log-normal distribution, $J_{\textrm{av}}$ shows compressed exponential behaviour, 
$J_{\textrm{av}}\propto \exp(\gamma\sigma^{n})$, where the exponent $n > 1$ is a non-universal parameter. We also study in detail field emission current for various electric field strength 
applied to systems with high density, characterised by Fermi energy, $E_{F} \gg \Phi$, $\Phi$ being the work function of the system as well as systems with low density characterised by 
$E_{F} \ll \Phi$.  
\end{abstract}
\maketitle
\section{Introduction}
Field emission is the process in which electrons from cold surfaces are emitted in the presence of applied strong electric field. This process should be compared 
agaist the thermo-ionic process in which electrons are emitted from hot metal surface. The field emission forms the back bone of modern semiconductor devices. This 
effect was first described by Fowler and Nordheim~\cite{FNpaper}. They considered quantum mechanical tunneling through a triangular potential energy (PE) barrier, 
created by the application of constant electric field. Much later Murphy and Good~\cite{MGpaper} (MG) introduced a more realistic PE barrier by taking into account the 
induced image charge formed in the presence of emitted electron. MG calculated barrier transmission coefficient under semi-classical WKB approximation. More recently, 
essentially an exact solution of the problem was obtained by Choy et. al.~\cite{ChoyJPCM2005}. Also, various cases including finite temperature (thermal emission), 
tunneling, curvature of the emission surface have been considered by various authors~\cite{CutlerSurfaceScience1993,ForbesUltramicro2001,EdgcombePRB2005,JensenAPL2006,
FischerJVacSciTech2013,ForbesJVacSciTech2013,KyritsakisProcRoySocA2015,HolgatePRAppl2017}

   To the best of our knowledge in all of those studies authors have considered constant \textit{local} work function. The work function of a material depends on the 
composition, structure, geometry, local charge distribution etc. of the emitting surface. Assumption of constant work function is only suitable for an atomistic smooth 
homogeneous surface. For surfaces with inhomogeneities over nano-scale (much smaller than the size of the collectors) assumption of constant work function is no longer 
valid. Also it has been shown~\cite{GamezJApplCrys2017} that in a system of nano-particles there is a distribution of the size of the nano-particles. Since the work 
function is more of a property of the surface we naturally can expect that the work function of nano-particles will also have a distribution. The actual microscopic model 
for work function distribution for a system of nano-particles is beyond the scope of this work. Interestingly, Gamez et. al.~\cite{GamezJApplCrys2017} using scanning 
tunneling microscope (STM) have measured the pair distribution function (PDF) for Pt nano-particles and they found that it follows log-normal distrubution.

    In this work, purely as a mathematical model, we choose Gaussian and log-normal distribution for the work function. We then study the field emision current averaged 
over work function distribution. The organization of the rest of the paper is as follows. In Sec. II we describe the mathematical formalism used to calculate field emission 
current. In Sec. III we describe the work function distribution used to calculate average current. In Sec. IV we present our results for both the case of Gaussian 
distribution and log-normal distribution. Finally in Sec. V we conclude.
\section{Mathematical formalism}
We closely follow and summarize the results obtained by Lopes et. al.~\cite{LopesCondMat,LopesPhysLett2020} for field-emission current density, $J$. In the standard FN-type MG theory 
the field emission current density is given by the well known expression~\cite{BiswasPhysPlasmas2017,HaugBook,ForbesAPL2006,ForbesJVacSciTech2010},  
\begin{eqnarray}
	J &=& \frac{e^3 \mathcal{E}^{2}}{16\pi^{2}\hbar\Phi}\frac{1}{t^{2}(y_{0})}
	\exp\left(-\frac{4}{3}\frac{\sqrt{2m}}{\hbar}\frac{\Phi^{3/2}}{e\mathcal{E}}v(y_{0})\right) 
	\label{Eq:Jtrans}
\end{eqnarray}	
where $e$ and $m$ are the charge and mass of electron, $\mathcal{E}$ is the strength of the applied electric field, $\Phi$ is the local work function of the emitting surface and 
\begin{eqnarray}
	v(y) = \left[\frac{1+\sqrt{1-y^{2}}}{2}\right]^{1/2}\left[E(\lambda)-\left(1-\sqrt{1-y^2}\right)K(\lambda)\right].
\end{eqnarray}
$K(\lambda)$ and $E(\lambda)$ are the complete elliptic integral of the first and second kind, with 
\begin{eqnarray}
	\lambda^{2} &=& \frac{2\sqrt{1-y^{2}}}{1+\sqrt{1-y^{2}}}
\end{eqnarray}
and 
\begin{eqnarray}
	y^{2} &=& \frac{e^{3}\mathcal{E}}{4\pi\epsilon_{0}(V_{0}-E)^{2}} < 1 .
\end{eqnarray}
It is important to mention that $V_{0}=\Phi+E_{F}$ is the height of the potential energy barrier ($E_{F}$ is the Fermi energy) and $E$ is the energy of the electron. 
Finally,
\begin{eqnarray}
	t(y_{0}) = \left[v(y)-\frac{2}{3}y\frac{dv}{dy}\right]_{y=y_{0}}
\end{eqnarray}
with 
\begin{eqnarray}
	y_{0}^{2} &=& \frac{e^{3}\mathcal{E}}{4\pi\epsilon_{0}\Phi^{2}}
	\label{eq:defn_y0}
\end{eqnarray}
From the relations above it is quite evident that calculation of transmission current requires evaluation of numerical integrals for complete elliptic integrals. 
Due to the singularities present in complete elliptic integrals it is very hard to extract meaningful results purely numerically~\cite{DolanPhysRev1953}. Under this 
circumstances we can consider series expansion for complete elliptic integrals~\cite{GradshteynBook} as follows
\begin{eqnarray}
	E(q) &=& 1+\frac{1}{2}\left[\ln\left(\frac{4}{q}\right)-\frac{1}{2}\right]q^{2}+\cdots \\
	K(q) &=& \ln\left(\frac{4}{q}\right)+\left[\ln\left(\frac{4}{q}\right)-1\right]\frac{q}{4} + \cdots
\end{eqnarray}
where 
\begin{eqnarray} 
	q &=& \sqrt{1-p^{2}} 
\end{eqnarray}
and $p^{2}=\frac{x_{2}-x_{1}}{x_{2}}$. $x_{1}$ and $x_{2}$ are the roots of the quadratic equation 
\begin{eqnarray}
	V_{0}-E -\frac{Ze^{2}}{4x} - e\mathcal{E}x = 0
\end{eqnarray}ls
A detailed calculation gives the following form for the field emission current density
\begin{eqnarray}
	J &=& \frac{e^{3}\mathcal{E}^{2}}{16\pi^{2}\hbar\Phi}\frac{\left[1-u(y_{0})\right]}{t^{2}(y_{0})}
	\exp\left[-\frac{4}{3}\sqrt{\frac{2m}{\hbar^{2}}}\frac{\Phi^{3/2}}{e\mathcal{E}}v(y_{0})\right]
	\label{Eq:Jmodified}
\end{eqnarray}
where
\begin{eqnarray}
	u(y_{0}) &=& \left[1+\frac{2\sqrt{2m\Phi}}{e\hbar}\frac{E_{F}}{\mathcal{E}}t(y_{0})\right]
	\exp\left[-\frac{2\sqrt{2m\Phi}}{e\hbar}\frac{E_{F}}{\mathcal{E}}t(y_{0})\right]
	\label{Eq:defn_uy0}
\end{eqnarray}
with
\begin{eqnarray}
	v(y_{0}) &=& 1 - \left[\frac{3}{8}Z'\ln\left(\frac{8}{\sqrt{Z'}}\right)+\frac{Z'}{16}\right]+\frac{3}{8}Z'y_{0}^{2} \ln(y_{0}) \nonumber \\ 
		 &+& \frac{Z'^{2}}{32}\left[1-\ln\left(\frac{8}{\sqrt{Z'}}\right)\right]y_{0}^{4}+\frac{Z'^{2}}{32}y_{0}^{4}\ln(y_{0}) + \cdots
\end{eqnarray}
and 
\begin{eqnarray}
	t(y_{0}) &=& v(y_{0}) -\frac{2}{3}y_{0}\frac{dv}{dy}(y_{0})
\end{eqnarray}
with $y_{0}$ given by Eq.~\ref{eq:defn_y0} and $Z'=1.179$ is an arbitrary constant obtained from suitable boundary condition~\cite{LopesCondMat}.

The expression for transmission current in Eq.~\ref{Eq:Jmodified} carries an interesting pre-exponential factor $[1-u(y_{0})]$ as compared to the standard form in 
Eq.~\ref{Eq:Jtrans}. This factor exhibits an explicit dependence on the external electric field ($\mathcal{E}$), work function ($\Phi$) and Fermi energy, $E_{F}$.

It is worth to mention that, from Eq.~\ref{Eq:Jmodified}, we can immediately investigate two important particular cases depending on the value of $E_{F}$. For large 
values of Fermi energies, i.e., $\frac{2\sqrt{2m}}{e\hbar}\frac{\Phi^{1/2}E_{F}}{\mathcal{E}} \gg 1$, or equivalently $E_{F} \gg \frac{e\hbar\mathcal{E}}{2\sqrt{2m\Phi}}$, 
it is easy to see that $u(y_{0})\approx 0$ in Eq.~\ref{Eq:defn_uy0} since the exponential dominates. So, in this limit we recover the standard FN-type MG equation. 

On the other hand, for the opposite regime, i.e., for very small Fermi energies, we have $E_{F} \ll \frac{e\hbar\mathcal{E}}{2\sqrt{2m\Phi}}$. In this limit we can 
expand the exponential in Eq.~\ref{Eq:defn_uy0} and get 
\begin{eqnarray}
	J =\frac{me}{2\pi^{2}\hbar^{3}} E_{F}^{2} \exp\left(-\frac{4}{3}\sqrt{\frac{2m}{\hbar^{2}}}\frac{\Phi^{3/2}}{e\mathcal{E}}v(y_{0})\right)
	\label{Eq:JtransSmallEF}
\end{eqnarray}
It is worth to emphasize that for small values of Fermi energies a very different expression, compared to the standard FN-type MG formula, was 
obtained~\cite{LopesCondMat} for the current density. Note that in this regime the pre-exponential factor in Eq.~\ref{Eq:JtransSmallEF} does not depend on 
either $t(y_{0})$ or the external electric field ($\mathcal{E}$), but it remains dependent on the Fermi energy ($E_{F}$).
\section{Work function distribution}
In the previous section we have summarized field emission current for a given \textit{local} work function $\Phi$. For a system of nano-particles or systems with 
inhomogeneity the work function will be different over the length scale of the size of the collector for emitted electrons. In such a situation we need to average 
the field emission current over the distribution of work function as follows :
\begin{eqnarray}
	J_{\textrm{av}} &=& \int P(\Phi) J(\Phi,E_{F}) d\phi
	\label{eq:Javerage}
\end{eqnarray}
where $P(\Phi)$ is the distribution of the work function, $\Phi$. We choose \textit{two} widely used distribution functions, namely (i) Gaussian and (ii) log-normal 
distribution. It is well known that systems with \textit{bulk disorder} follows Gaussian distribution and the work function distribution function, $P(\Phi)$, is given by
\begin{eqnarray}
	P_{N}(\Phi) &=& \frac{1}{\sigma\sqrt{2\pi}} \exp\left[-\frac{(\Phi-\Phi_{0})^{2}}{2\sigma^{2}}\right],
	\label{eq:NormalDist}
\end{eqnarray}
where $\sigma$ is the \textit{standard deviation} of the distribution and $\Phi_{0}$ is the known \textit{bulk} value for a given material.

We also use log-normal distribution for the work function :
\begin{eqnarray}
	P_{LN}(\Phi) &=& \frac{1}{\Phi\sigma\sqrt{2\pi}} \exp\left[-\frac{(\ln \Phi - \mu)^{2}}{2\sigma^{2}}\right]
	\label{eq:LNdist}
\end{eqnarray}
It is important to mention that we were inspired by the experimental result of Gamez et. al.~\cite{GamezJApplCrys2017}. They showed that for a system of Pt nano-particles 
the pair distribution function (PDF) for the radius of nano-particles follows log-normal distribution 
\begin{eqnarray}
	F(r) = \frac{1}{rs\sqrt{2\pi}}\exp\left[-\frac{(\ln r-\mu)^2}{2s^2}\right]
\end{eqnarray}
with 
\begin{eqnarray}
	s^{2} &=& \ln\left[\left(\frac{\textrm{P}_{\textrm{sig}}}{\textrm{P}_{\textrm{size}}}\right)+1\right]\\
	\mu &=& \ln\left(\textrm{P}_{\textrm{size}}\right)-\frac{s^{2}}{2}
\end{eqnarray}
where $\textrm{P}_{\textrm{size}}$ and $\textrm{P}_{\textrm{sig}}$ are the average particle diameter and standard deviation, respectively. Since work function crucially depends on the 
surface properties of a system it may follow log-normal distribution as in the case of \textit{surface disorder}. 

Various statistical properties of this distribution are summarized in the table~\ref{tab:table1} :
\begin{table}[h!]
	\begin{center}
		\caption{Statistical properties of log-normal distribution}
		\label{tab:table1}
	\begin{tabular}{|l|l|}
		\hline
		Mean & $e^{\mu+\frac{\sigma^2}{2}}$\\
		\hline
		Variance & $\nu = e^{2\mu+\sigma^2}\left[e^{\sigma^2}-1\right]$\\
		\hline
		Standard deviation & $s=\sqrt{\nu}$\\
		\hline
		Median & $M=e^{\mu}$\\
		\hline
		Mode   & $\Gamma = e^{\mu-\sigma^2}$\\
		\hline
		Skewness & $\Sigma = (e^{\sigma^2}+2)\sqrt{e^{\sigma^2}-1}$\\
		\hline
		Kurtosis & $\kappa = e^{4\sigma^{2}}+2e^{3\sigma^2}+3e^{2\sigma^2}-3$\\
                \hline
        \end{tabular}
	\end{center}
\end{table}
\section{Results}
In this section we show our results for current density averaged over two choice of probability distributions as have been discussed in the previous section.  
\subsection{Gaussian work function distribution}
\begin{figure}[htbp]
\begin{center}
	\subfloat{\includegraphics[height=22ex]{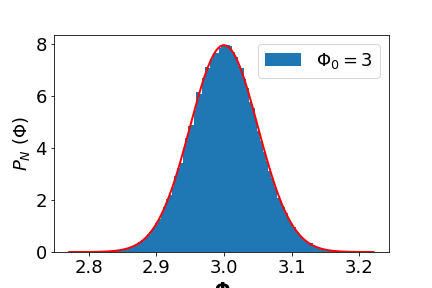}}
	\subfloat{\includegraphics[height=22ex]{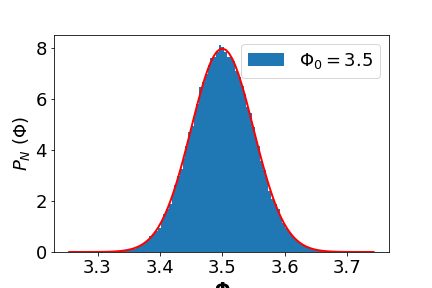}}\\
	\subfloat{\includegraphics[height=22ex]{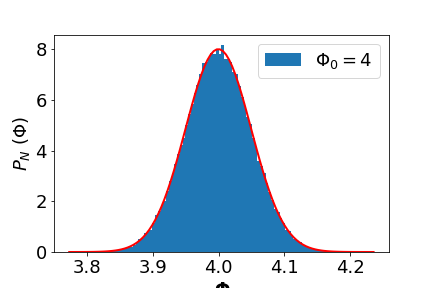}}
	\subfloat{\includegraphics[height=22ex]{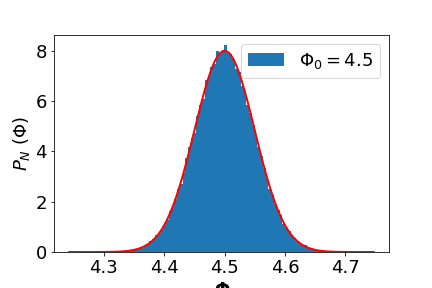}}
\end{center}
	\caption{(Color online) Histogram plot of work function, $\Phi$, sampled over Gaussian distribution for 
	various average work function, $\Phi_{0}$. Solid line shows fitting to Gaussian distribution and we have chosen 
	$\sigma=0.05$ for all the cases.}
	\label{Fig:GaussianDist}	
\end{figure}

We first consider the case of Gaussian distribution for the work function. In Fig.~\ref{Fig:GaussianDist} we show the histogram plot of the work function, sampled over 
Gaussian distribution for \textit{four} choices of \textit{bulk} work function $\Phi_{0}=3.0, 3.5, 4.0$ and $4.5$ eV with $\sigma=0.05$. In each case we also fit the histogram 
plot to Gaussian distribution. From this fit we can see that our choice of random variables for $\Phi$ are well sampled over Gaussian distribution.
\subsubsection{Case with $\Phi \ll E_{F}$}
\begin{figure}[htbp]
\centering
\includegraphics[clip,scale=0.32]{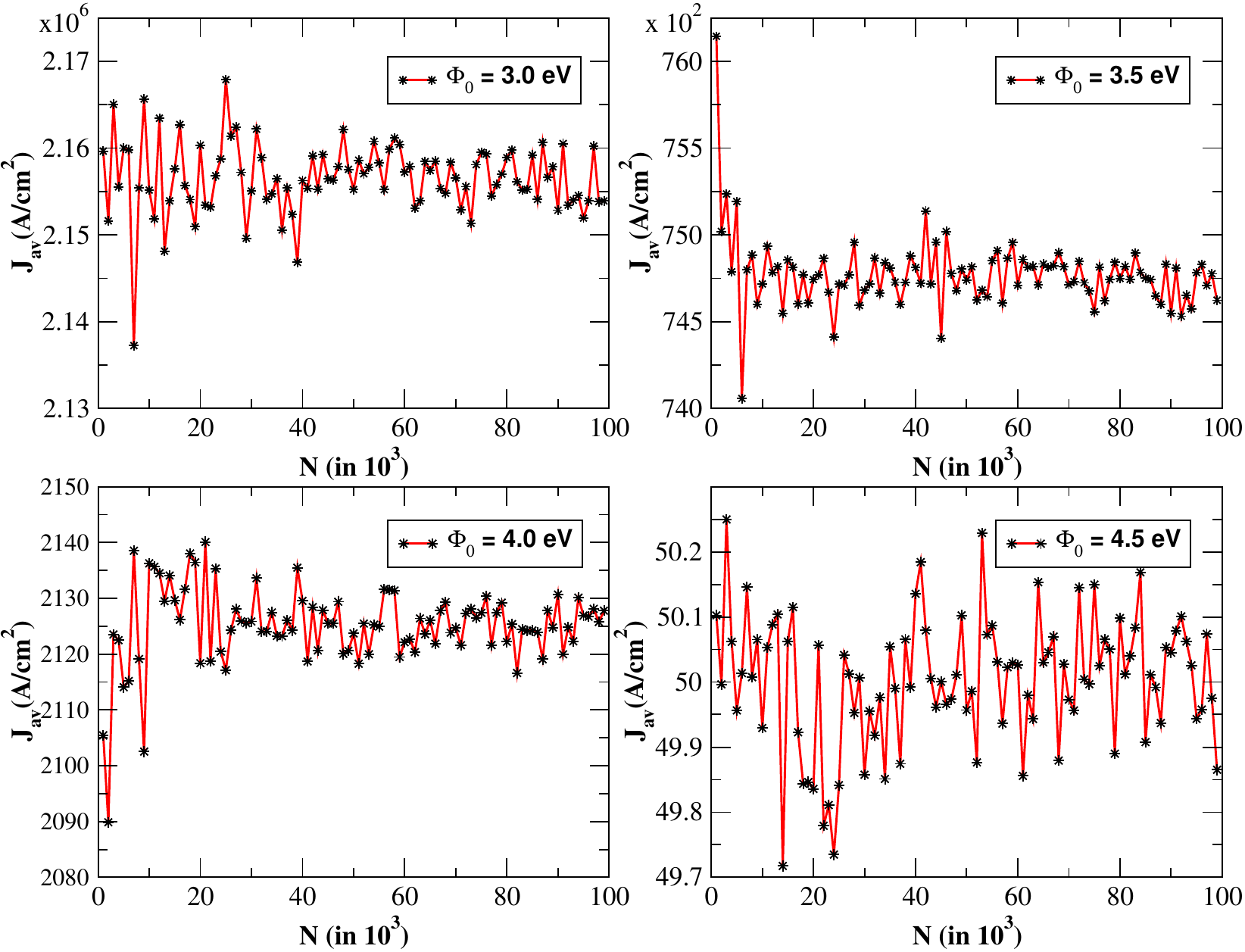}
\caption{(Color online) Average current, $J_{\textrm{av}}$, as a function of number of random variables, N chosen from Gaussian distribution for 
$\Phi_{0} \ll E_{F}$.}
\label{Fig:JvsNEF10.0Norm}
\end{figure}

We now consider the case with $E_{F}=10$ eV and $\Phi_{0} =$ 3.0, 3.5, 4.0 and 4.5 eV. Since $E_{F}\propto n^{2/3}$, $n$ being the density of electrons, this will 
correspond to high density limit. In Fig.~\ref{Fig:JvsNEF10.0Norm} we show the current density averaged over various number of random variables, $N$, for $\Phi$. For 
$N < 10^{4}$ there is significant fluctuations of $J_{\textrm{av}}$ but for $N > 4\times10^{4}$, $J_{\textrm{av}}$ show small fluctuations over mean value. We choose 
$N = 10^{5}$ for calculation of $J_{\textrm{av}}$ for all the cases and $J_{\textrm{av}}$ will be free of \textit{statistical errors.} 

As shown in Fig.~\ref{Fig:JvsNEF10.0Norm}, $J_{\textrm{av}}$ varries over several orders of magnitude as we varry $\Phi_{0}$ from 3.0 eV to 4.5 eV. So in order to understand 
role of work function distribution on field emission current we need to study the dimensionless scaled quantity $J_{\textrm{av}}/J_{0}$, where $J_{0}$ is the field emission 
current for \textit{bulk} value of the work function $\Phi_{0}$ and can be calculated from  Eq.~\ref{Eq:Jtrans} with $\Phi = \Phi_{0}$.

\begin{figure}[htbp]
\centering
	\includegraphics[clip,scale=0.32]{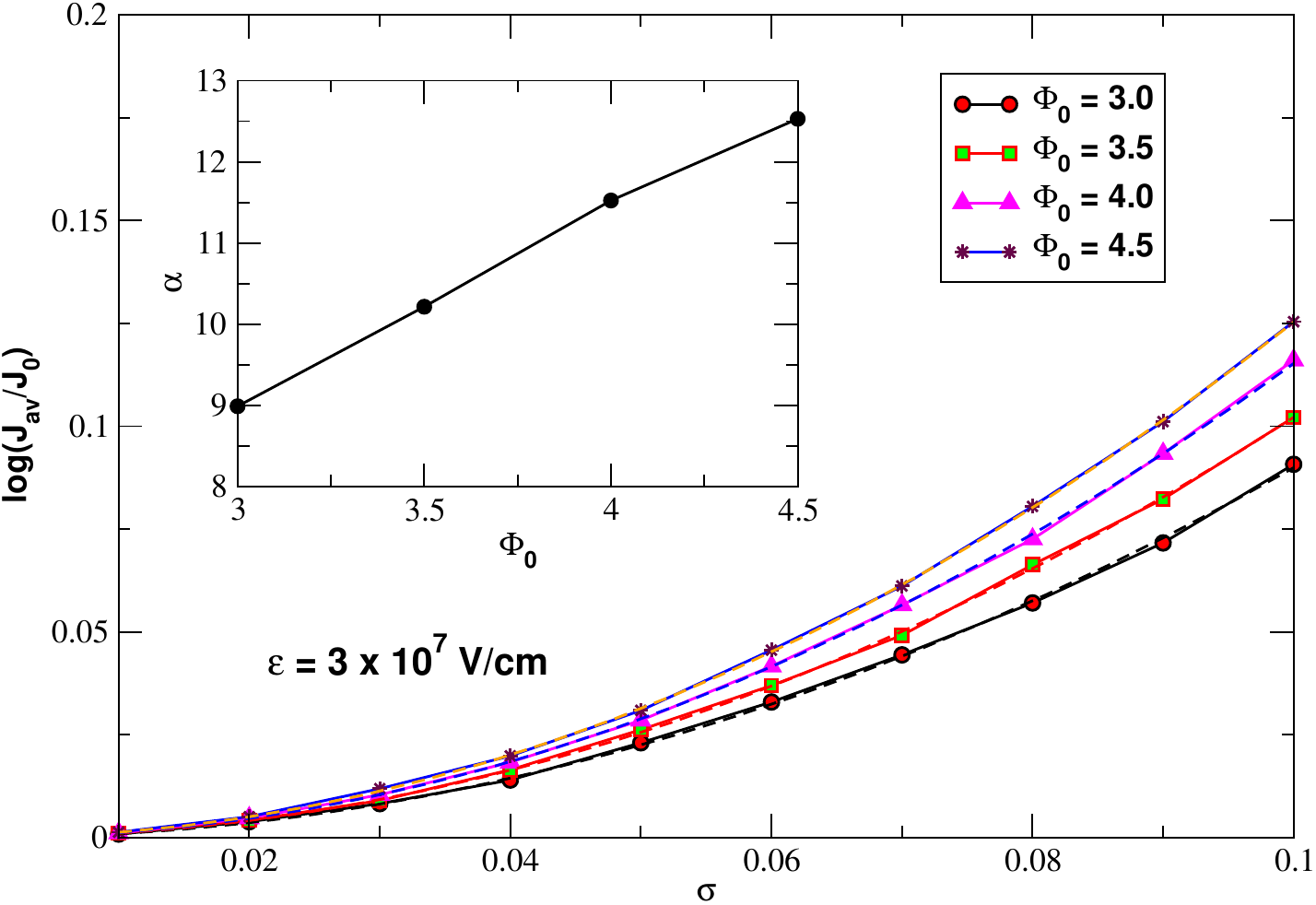}
	\caption{(Color online) Logarithm of the scaled average current, $J_{\textrm{av}}/J_{0}$, as a function of the width, $\sigma$, of the work function distribution for 
	 various average work function, $\Phi_{0} \ll E_{F}$. Solid lines with symbols are the numerical data and the dashed lines represents fitting function 
	 $f(\sigma)=\alpha\sigma^{2}$. Inset : variation of the fitting parameter $\alpha$ with $\Phi_{0}$. The strength of the electric field, 
	 $\mathcal{E} = 3 \times 10^{7}\textrm{V/cm}$, and Fermi energy $E_{F}=10.0$ eV for all the cases.}
	\label{Fig:JavVsSigmaMuEF10.0}
\end{figure}

In Fig.~\ref{Fig:JavVsSigmaMuEF10.0} we show, $\log(J_{\textrm{av}}/J_{0})$, as a function of the width of the work function distribution, $\sigma$ for a given applied electric 
field, $\mathcal{E} = 3 \times 10^{7}\textrm{V/cm}$. We choose $0.01 \leq \sigma \leq 0.1$. Since the full width at half maxima (FWHM) for a Gaussian distribution is $2.35\sigma$ 
and more than $95\%$ integrated weight is within a width $4\sigma$ (between $-2\sigma$ and $+2\sigma$) our choice of $\sigma$ ensures that the deviation of $\Phi$ from its 
bulk value $\Phi_{0}$ is less than $10\%$. The first noticeable feature in Fig.~\ref{Fig:JavVsSigmaMuEF10.0} is that the current density, $J_{\textrm{av}}$, averaged over work 
function distribution increases monotonically with the width of the distribution, $\sigma$. This is because the transport current, $J(\Phi) \propto \exp(-\zeta \Phi^{3/2})$ where 
$\zeta = \frac{4}{3}\sqrt{\frac{2m}{\hbar^{2}}}\frac{v(y_{0})}{e\mathcal{E}}$ and the averaging over work function distribution leads to exploration of regions with lower barrier 
height hence increase of tunneling current. 

Another noticeable feature in Fig.~\ref{Fig:JavVsSigmaMuEF10.0} is that the logarithm of the scaled average current $\log(J_{\textrm{av}}/J_{0})$ can be fitted well with the 
functional form $f(\sigma) = \alpha\sigma^{2}$, where $\alpha$ is a fitting parameter. This clearly shows that $J_{\textrm{av}}$ follows Gaussian behaviour 
$J_{\textrm{av}} = J_{0}\exp(\alpha\sigma^{2})$. The fitting parameter $\alpha$ in general depends on $\Phi_{0},\mathcal{E}$ and $E_{F}$. The dependence of $\alpha$ with $\Phi_{0}$ is 
nearly linear as shown in the inset of Fig.~\ref{Fig:JavVsSigmaMuEF10.0}.  

\begin{figure}[htbp]
\centering
	\includegraphics[clip,scale=0.32]{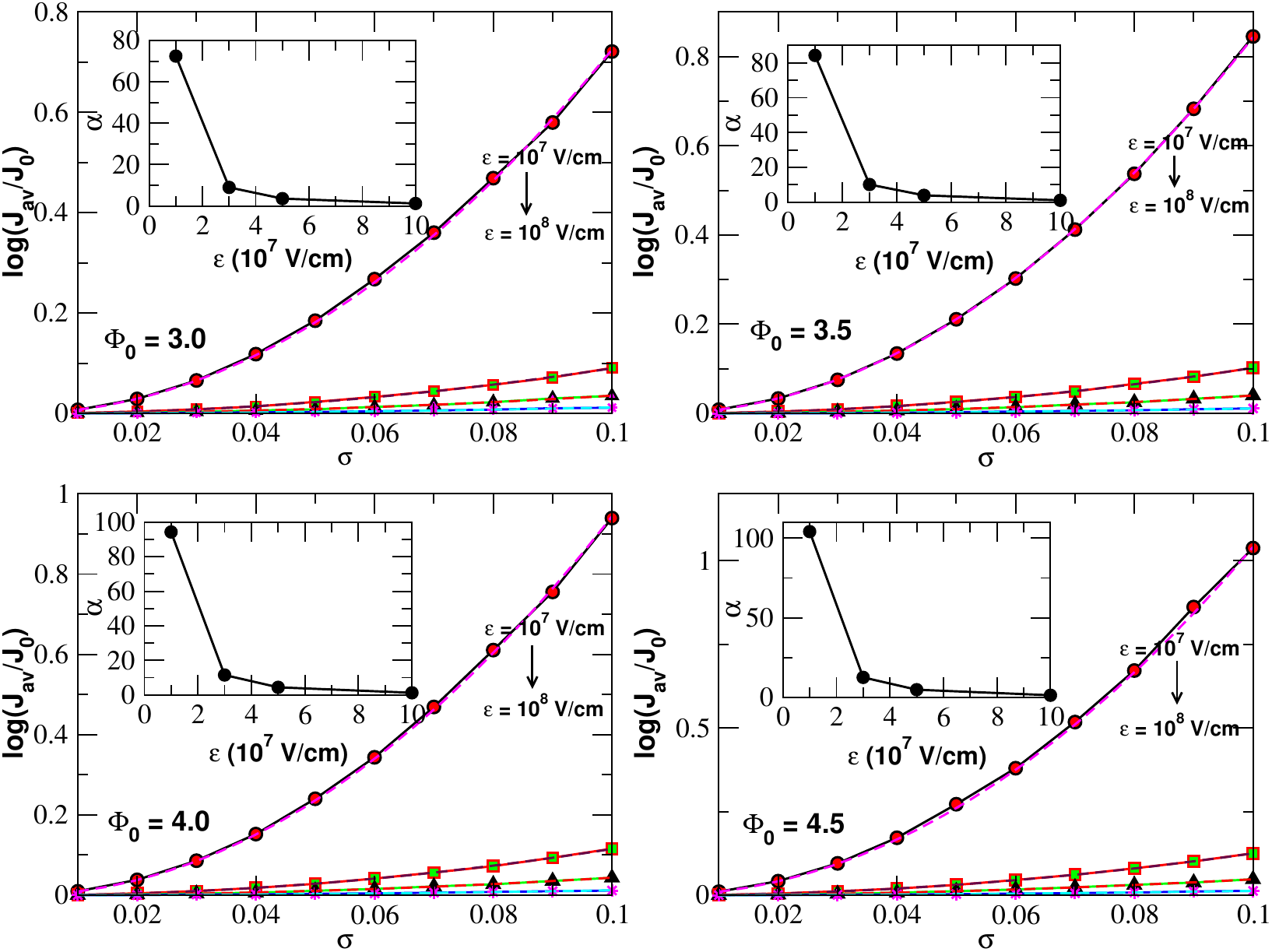}
	\caption{(Color online)Panel plot : Logarithm of scaled average current, $J_{\textrm{av}}/J_{0}$, as a fuction of the width, $\sigma$ of the work function distribution for various field 
	strength, $\mathcal{E}$. Solid lines with symbols are numerical data and the dashed lines represent fitting function $f(\sigma)=\alpha\sigma^{2}$. Inset of each plot shows variation of 
	$\alpha$ with field strength, $\mathcal{E}$. The value of $\Phi_{0}$ is indicated in each plot and $E_{F}=10.0$ eV for all the plots.}
	\label{Fig:PanelPlotJavByJ0vsFEF10.0}
\end{figure}

In Fig.~\ref{Fig:PanelPlotJavByJ0vsFEF10.0} we show the behaviour of $\log(J_{\textrm{av}}/J_{0})$ for various field strength $\mathcal{E}$ as well as $\Phi_{0}$. 
For $\mathcal{E}=10^{7}$ V/cm the increase of $J_{\textrm{av}}$ is more than $e$ times over the bulk value $J_{0}$ for $\sigma=0.1$. This can be explained by the fact that the 
tunneling current $J \propto \frac{\mathcal{E}^{2}}{\Phi}\exp(-\zeta \Phi^{3/2}/\mathcal{E})$. With increasing field strength the tunneling current dramatically increases mainly 
due to its $\exp(-1/\mathcal{E})$ dependence. Averaging over work function distribution increases tunneling current as explained earlier. However this effect becomes sub-dominant 
for larger field strengths as the tunneling current increses over several orders of magnitude for just one order of magnitude increase of $\mathcal{E}$. The behaviour is similar 
for all the four choices of $\Phi_{0}$. In the inset of each plot in Fig.~\ref{Fig:PanelPlotJavByJ0vsFEF10.0} we have shown the dependence of fitting parameter $\alpha$ with the 
field strength $\mathcal{E}$. As can be clearly seen $\alpha$ very strongly depends on $\mathcal{E}$.

\subsubsection{Case with $\Phi \gg E_{F}$}
\begin{figure}[htbp]
\centering	
\includegraphics[clip,scale=0.32]{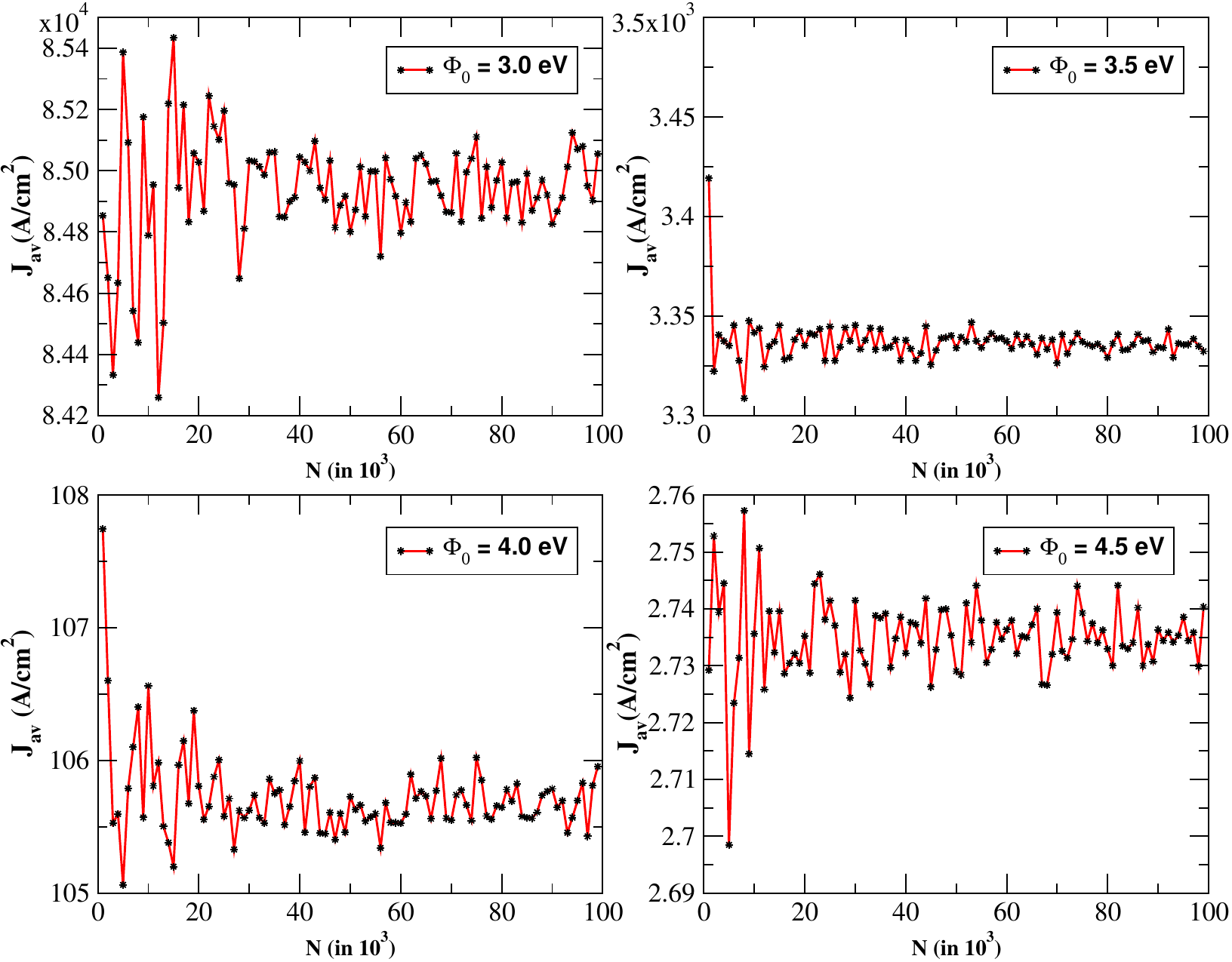}
\caption{(Color online)Average current, $J_{\textrm{av}}$, as a function of number of random variables, N, chosen from Gaussian distribution for $\Phi_{0} \gg E_{F}$.}
\label{Fig:JvsNEF0.05Norm}
\end{figure}

Next we consider the case with $E_{F}=0.05$ eV and $\Phi_{0} = 3.0, 3.5, 4.0$ and 4.5 eV . This will correspond to systems with low density as $E_{F}\propto n^{2/3}$. In this case the tunneling 
current does not depend on the prefactor $\frac{e^{3}\mathcal{E}^{2}}{16\pi^{2}\hbar\Phi}\frac{[1-u(y_{0})]}{t^{2}(y_{0})}$ but still depends on the standard exponential factor 
$\exp(-\zeta \Phi^{3/2}/\mathcal{E})$. In Fig.~\ref{Fig:JvsNEF0.05Norm} we show $J_{\textrm{av}}$ as a function of $N$, the number of random variables for averaging over work function distribution 
for $\Phi_{0} = 3.0, 3.5, 4.0$ and 4.5 eV. As can be clearly seen $J_{\textrm{av}}$ is atleast one order of magnitude smaller than the case with $\Phi \ll E_{F}$. This is mainly due to the fact that 
the prefactor $\frac{me}{2\pi^{2}\hbar^{3}} E_{F}^{2}$ for tunneling current $J$ in this case is much smaller than the prefactor 
$\frac{e^{3}\mathcal{E}^{2}}{16\pi^{2}\hbar\Phi}\frac{[1-u(y_{0})]}{t^{2}(y_{0})}$ corresponding to the case with $\Phi_{0} \ll E_{F}$. As in the earlier case $J_{\textrm{av}}$ shows significant 
fluctuation for $N < 10^{4}$. We choose $N=10^{5}$ for various $\Phi_{0}$ and $\mathcal{E}$. 

\begin{figure}[htbp]
\centering
	\includegraphics[clip,scale=0.32]{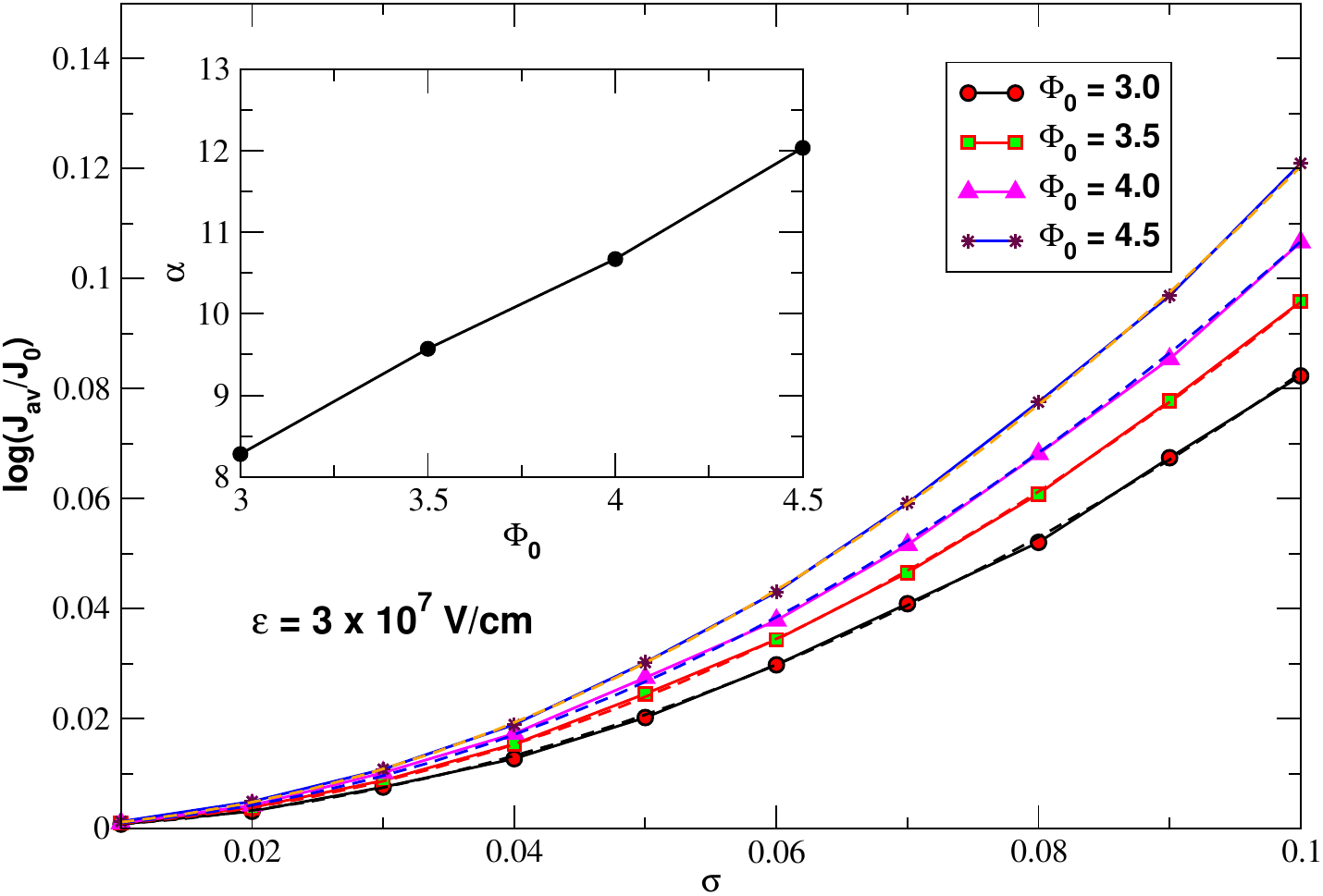}
	\caption{(Color online) Logarithm of scaled average current, $J_{\textrm{av}}/J_{0}$, as a function of the width, $\sigma$, of the Gaussian work function distribution 
	for various average work function, $\Phi_{0} \gg E_{F}$. Solid lines with symbols are numerical data and dashed lines represents fitting function 
	$f(\sigma) = \alpha\sigma^{2}$. Inset : variation of the fitting parameter $\alpha$ with $\Phi_{0}$. The strength of the electric field $\epsilon = 3\times 10^{7} \textrm{V/cm}$ and 
	$E_{F}=0.05$ eV.}
	\label{Fig:PanelPlotJavByJ0vsMuEF0.05}
\end{figure}

In Fig.~\ref{Fig:PanelPlotJavByJ0vsMuEF0.05} we show $\log(J_{\textrm{av}}/J_{0})$ as a function of the width, $\sigma$, of the Gaussian work function distribution for a given applied 
electric field strength $\mathcal{E}= 3 \times 10^{7}\textrm{V/cm}$. As in the case with $\Phi_{0} \ll E_{F}$, $J_{\textrm{av}}$ monotonically increases with $\sigma$. But the increase is 
slower than the case with $\Phi_{0} \ll E_{F}$. This is mainly because of the absence of $\Phi$ dependent prefactor in the expression for $J$. Most intrestingly, as in the earlier case with 
$\Phi_{0} \ll E_{F}$, $\log(J_{\textrm{av}}/J_{0})$ can be well fitted with $f(\sigma) = \alpha\sigma^{2}$. The fitting parameter $\alpha$ linearly depends on $\Phi_{0}$.

\begin{figure}[htbp]
\centering
\includegraphics[clip,scale=0.32]{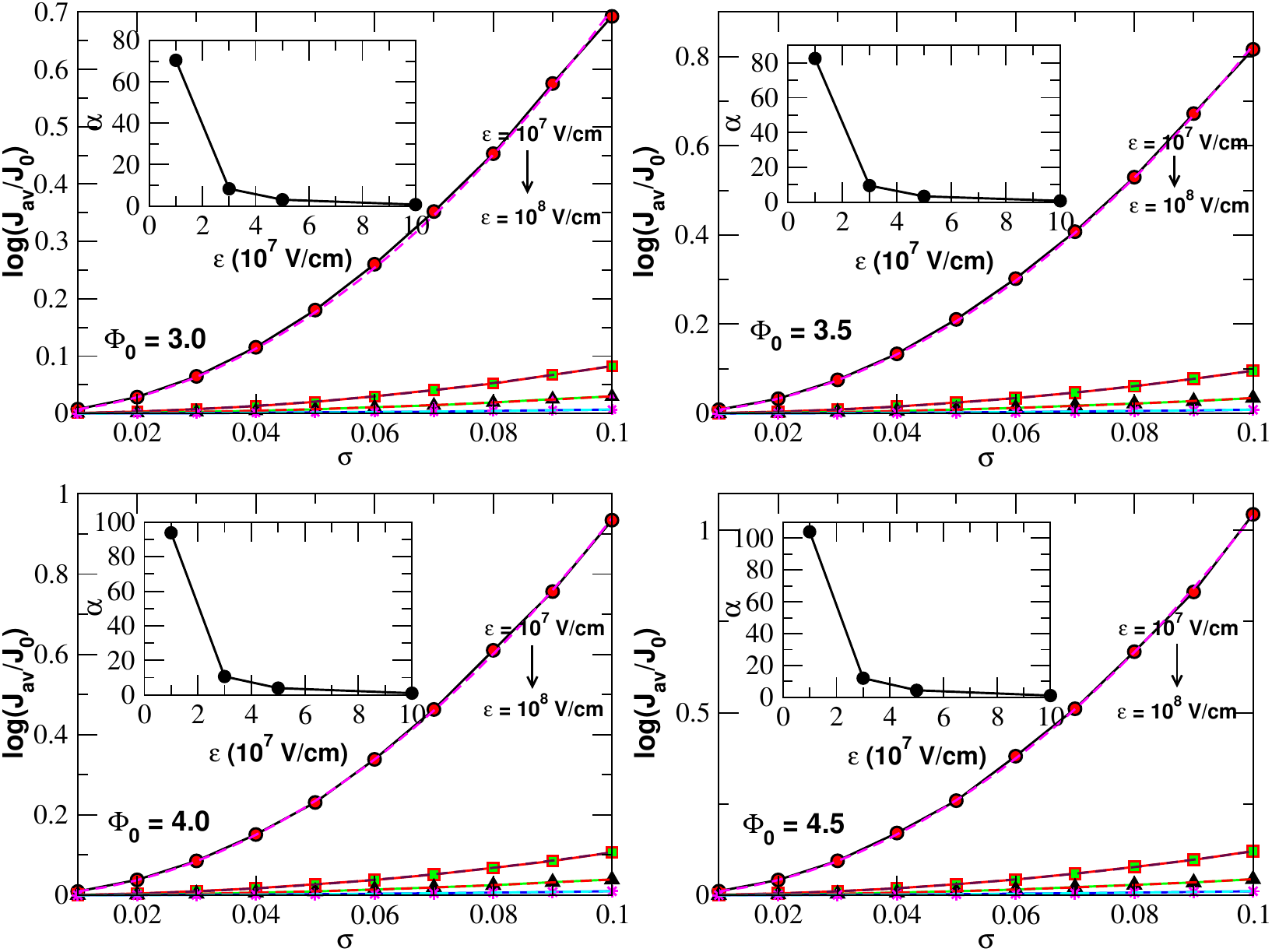}
\caption{(Color online) Panel plot : Logarithm of the scaled average current, $J_{\textrm{av}}/J_{0}$, as a function of the width, $\sigma$ of the Gaussian work function distribution for 
	various electric field strength, $\mathcal{E}$. Solid lines with symbols are numerical data and the dashed lines are fitting function $f(\sigma)=\alpha\sigma^{2}$. Inset of each plot shows 
	the variation of $\alpha$ with field strength, $\mathcal{E}$. The value of $\Phi_{0} \gg E_{F}$ is indicated in each plot and $E_{F}=0.05$ eV for all the plots.}
\label{Fig:PanelPlotJavByJ0VsFMuEF0.05}
\end{figure}

In Fig.~\ref{Fig:PanelPlotJavByJ0VsFMuEF0.05} we show the dependence of $\log(J_{\textrm{av}}/J_{0})$ with $\sigma$ for various field strength $\mathcal{E}$ and $\Phi_{0}$. As in the case with 
$\Phi_{0} \ll E_{F}$, effect of work function distribution is strongest for the weakest field $\mathcal{E} = 10^{7}$ V/cm. The tunneling current for $\mathcal{E}=10^{7}$ V/cm is very small 
(~$10^{-10} \textrm{A/cm}^{2}$) and very small change (lowering) of potential barrier can have much larger effect. With increasing field strength by one order of 
magnitude changes the tunneling current by several orders of magnitude which largely cancels the effect of work function distribution. However for each case $\log(J_{\textrm{av}}/J_{0})$ can be 
fitted well with $f(\sigma) = \alpha\sigma^{2}$. The fitting parameter $\alpha$ strongly depends on $\mathcal{E}$ as shown in the inset of each plot. 

\subsection{Log-normal work function distribution}
\begin{figure}[htbp]
\begin{center}
	\subfloat{\includegraphics[height=22ex]{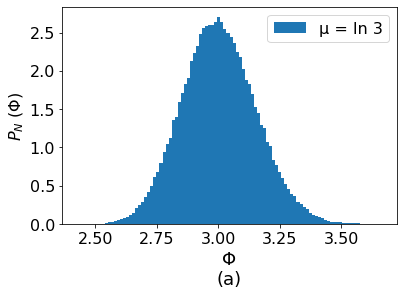}}
	\subfloat{\includegraphics[height=22ex]{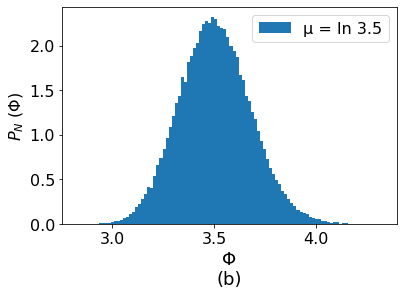}}\\
	\subfloat{\includegraphics[height=22ex]{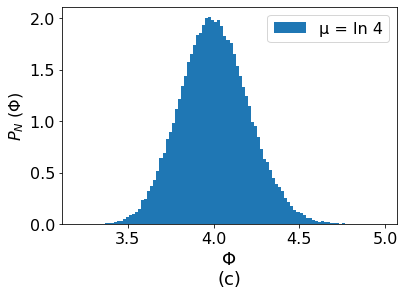}}
	\subfloat{\includegraphics[height=22ex]{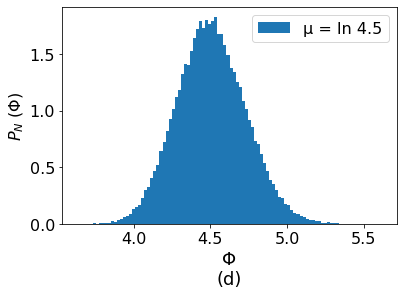}}
\end{center}
	\caption{(Color online) Histogram plot of work function, $\Phi$, sampled over log normal distribution for 
	various $\mu$ (see text). We have chosen $\sigma=0.05$ for all the figures.}
	\label{Fig:LogNormalDist}	
\end{figure}

Next we consider log-normal distribution for the work function. In Fig.~\ref{Fig:LogNormalDist} we show the histogram plot of the work function, sampled over log-normal distribution for 
\textit{four} choice of median work function $M\equiv e^{\mu} = 3.0, 3.5, 4.0$ and 4.5 eV with $\sigma=0.05$. From this plot we can see that our choice of random variables for $\Phi$ are well 
sampled over the log-normal distribution.

\subsubsection{Case with $\Phi \ll E_{F}$}
\begin{figure}[htbp]
\centering	
\includegraphics[clip,scale=0.32]{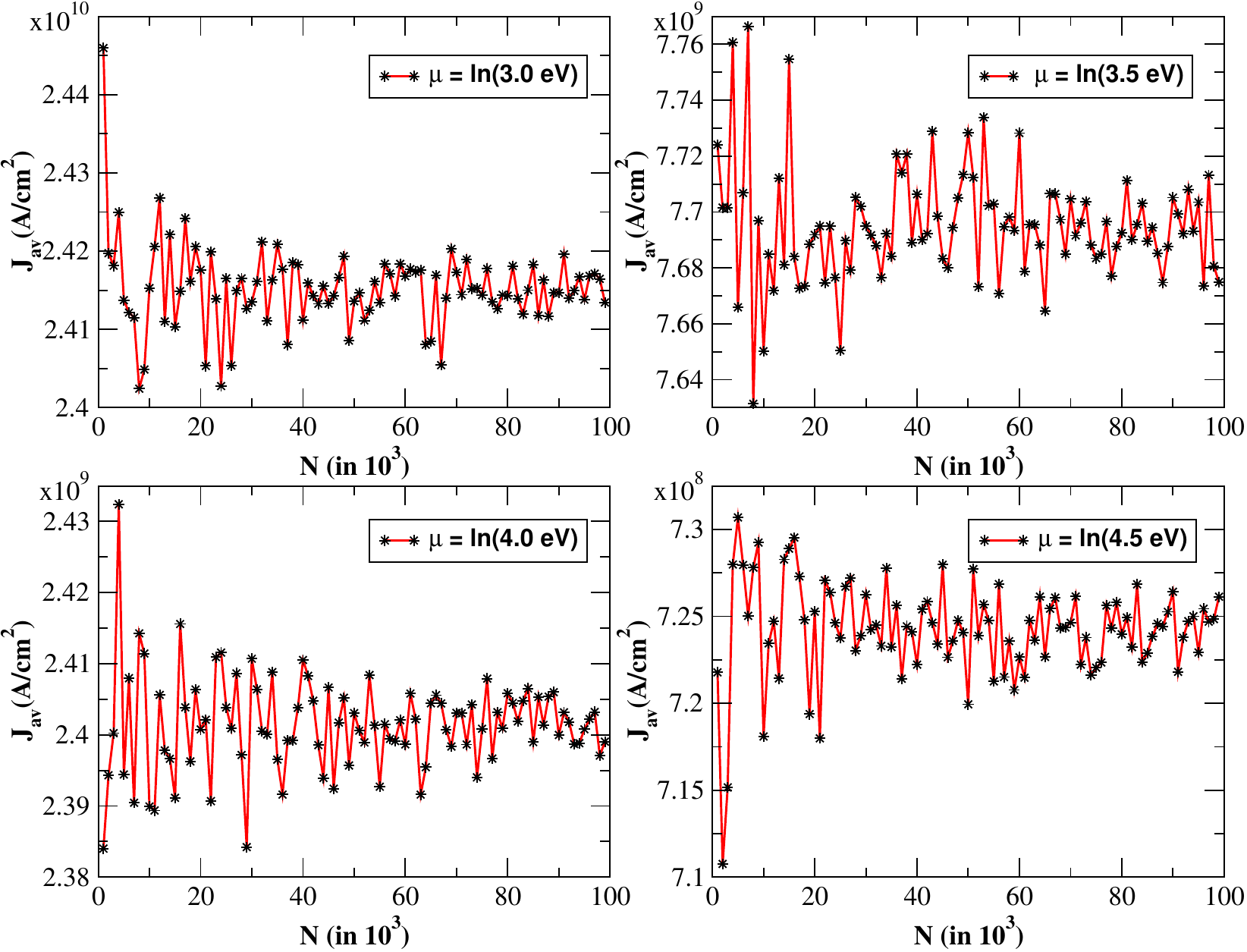}
\caption{(Color online)Average current, $J_{\textrm{av}}$, as a function of number of random variables, N, chosen from Log-normal distribution for 
$\Phi \ll E_{F}$.}
\label{Fig:JvsNEF10.0LogNorm}
\end{figure}

First we consider the familiar case with $\Phi \ll E_{F}$. We choose $E_{F}=10$ eV and $\mu = \ln 3.0, \ln 3.5, \ln 4.0$ and $\ln 4.5$. This will correspond to median value $M=3.0, 3.5, 4.0$ and 
4.5 eV. In Fig.~\ref{Fig:JvsNEF10.0LogNorm} we show $J_{\textrm{av}}$ as a function of number of random variables for $\Phi$, chosen from log-normal distribution. The current density is 
several orders of magnitude higher than the case of Gaussian distribution. Also the fluctuations are higher than the case of Gaussian distribution. As in the earlier cases, we choose $N=10^{5}$ to 
get good statistical measure. 

\begin{figure}[htbp]
\centering
\includegraphics[clip,scale=0.32]{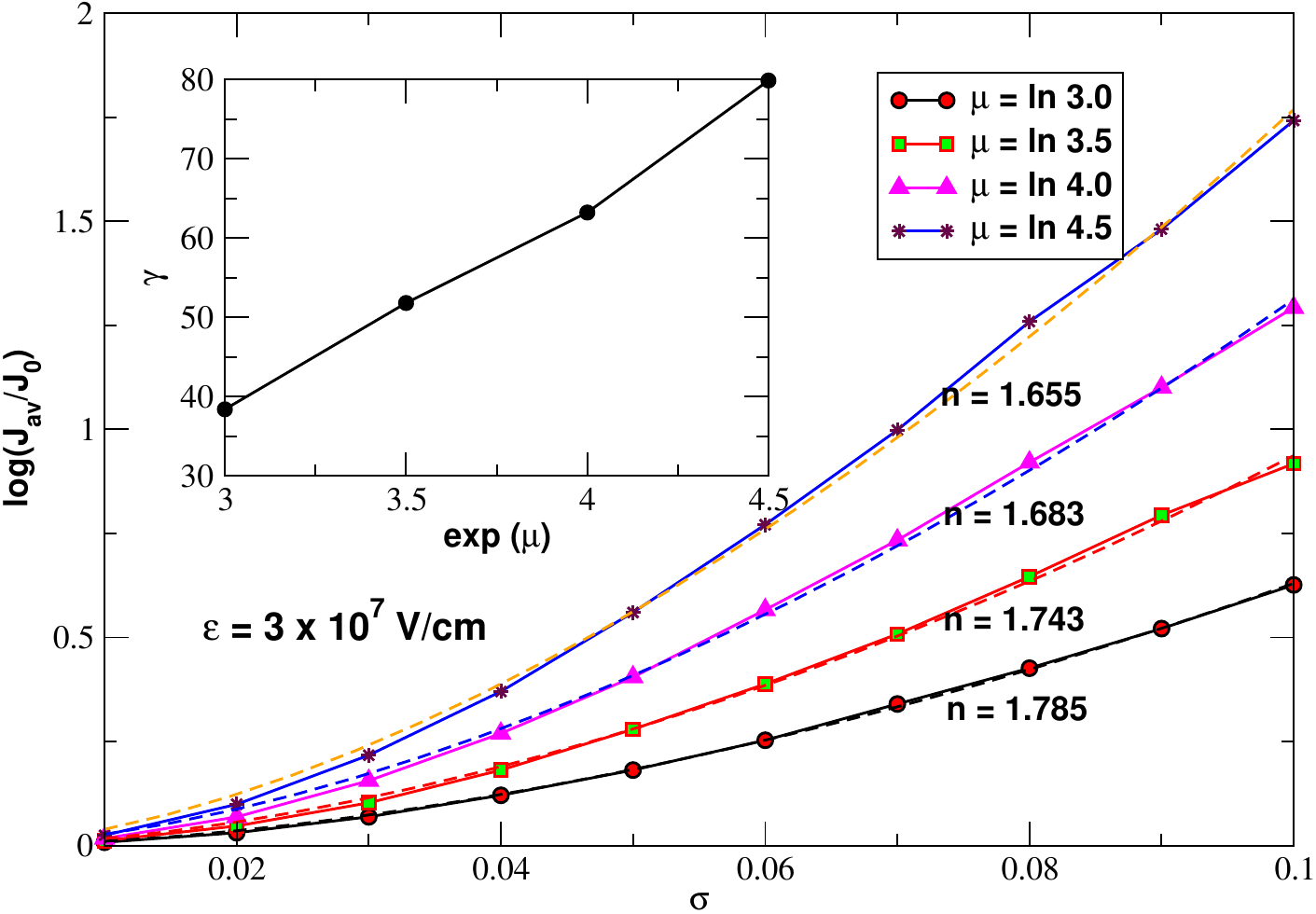}
\caption{(Color online) Logarithm of scaled average current, $J_{\textrm{av}}/J_{0}$, as a function of the width, $\sigma$, of the log-normal work function distribution for $\Phi \ll E_{F}$. 
Solid lines with symbols are numerical data and the dashed lines are fitting to $g(\sigma) = \gamma\sigma^{n}$, where $n > 1$ is a non-universal exponent characterizing compressed exponential 
behaviour of $J_{\textrm{av}}$. Inset : variation of fitting parameter $\gamma$ with the median of the distribution $M=\exp(\mu)$. The value of the exponents $n$ for each case is shown on the 
figure and the value of the field strength $\epsilon=3\times 10^{7} \textrm{V/cm}$ and $E_{F}=10$ eV.}
\label{Fig:JavVsSigmaMuEF10.0LogNorm}
\end{figure}

In Fig.~\ref{Fig:JavVsSigmaMuEF10.0LogNorm} we show $\log(J_{\textrm{av}}/J_{0})$ as a function of the width ($\sigma$) of the log-normal work function distribution for $M = 3.0, 3.5, 4.0$ and 
4.5 eV and a given field strength $\mathcal{E}=3\times 10^{7}$ V/cm, respectively. As in the case of Gaussian distribution, $J_{\textrm{av}}$ increases monotonically as a function of the width  
of the distribution, $\sigma$. However the increase in this case is much stronger - more than 10 times $J_{0}$ for $\sigma=0.1$. This strong increase can be attributed to the skewness of 
distribution function. There is significantly higher probability of lowering of effective potential barrier than the Gaussian distribution with no skewness. As a result, the current density 
$J \propto \exp(-\zeta\Phi^{3/2}/\mathcal{E})$ will show exponential increase with the width of the distribution. As shown in Fig.~\ref{Fig:JavVsSigmaMuEF10.0LogNorm}, $\log(J_{\textrm{av}}/J_{0})$ 
can be fitted well with the function $g(\sigma)=\gamma \sigma^{n}$ with non-universal exponent $n > 1$. The fitting parameter $\gamma$ increases linearly with the mean of the distribution. 

\begin{figure}[htbp]
\centering
\includegraphics[clip,scale=0.32]{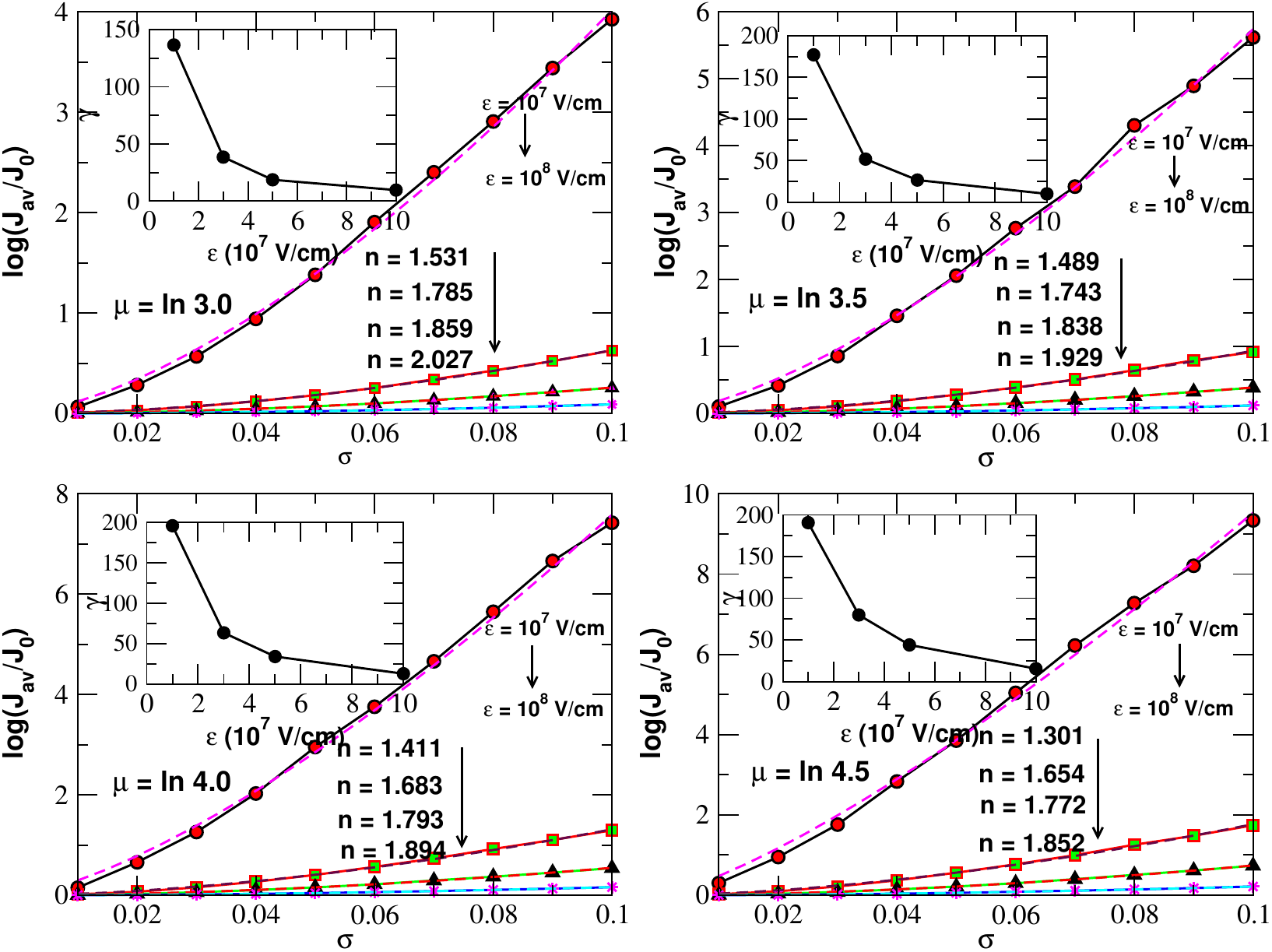}
\caption{(Color online) Panel plot : Logarithm of scaled average current, $J_{\textrm{av}}/J_{0}$, as a function of the width, $\sigma$, of the log-normal work function distribution for 
various applied field strength, $\mathcal{E}$. $\mathcal{E}$ varies from $10^{7} \textrm{V/cm}$ (top most curve) to $10^{8} \textrm{V/cm}$ (bottom most curve) in each panel. Solid lines with 
symbols are numerical data and dashed lines represent fitting $g(\sigma) = \gamma\sigma^{n}$, where $n > 1$ is non-universal exponent. Inset of each plot shows variation of $\gamma$ 
with $\mathcal{E}$ and the variation $n$ is indicated in each plot. $\mu$ is shown in each plot and $E_{F}=10.0$ eV for all the plots.}
\label{Fig:PanelPlotJavByJ0vsFEF10.0LogNorm}
\end{figure}

In Fig.~\ref{Fig:PanelPlotJavByJ0vsFEF10.0LogNorm} we show $\log(J_{\textrm{av}}/J_{0})$ as a function of the width, $\sigma$, of the log-normal work function distribution function for various 
field strength $\mathcal{E}$ varying over one order of magnitude - from $10^{7} \textrm{V/cm}$ to $10^{8} \textrm{V/cm}$. For the lowest field $\mathcal{E}=10^{7}\textrm{V/cm}$, $J_{\textrm{av}}$ 
increases by several orders of magnitude over $J_{0}$. This is mainly because of the fact that low field tunneling current is very small and the effect of averaging over work function distribution 
together with the skewness of the distribution increases the tunneling current drastically. Also the rise is near exponential as $n \sim 1$. With increasing $\mathcal{E}$ the tunneling current 
increases and the effect of work function distribution gets largely suppressed. The fitting parameter $\gamma$ strongly depends on the field strength $\mathcal{E}$.

\subsubsection{Case with $\Phi \gg E_{F}$}
\begin{figure}[htbp]
\centering	
\includegraphics[clip,scale=0.32]{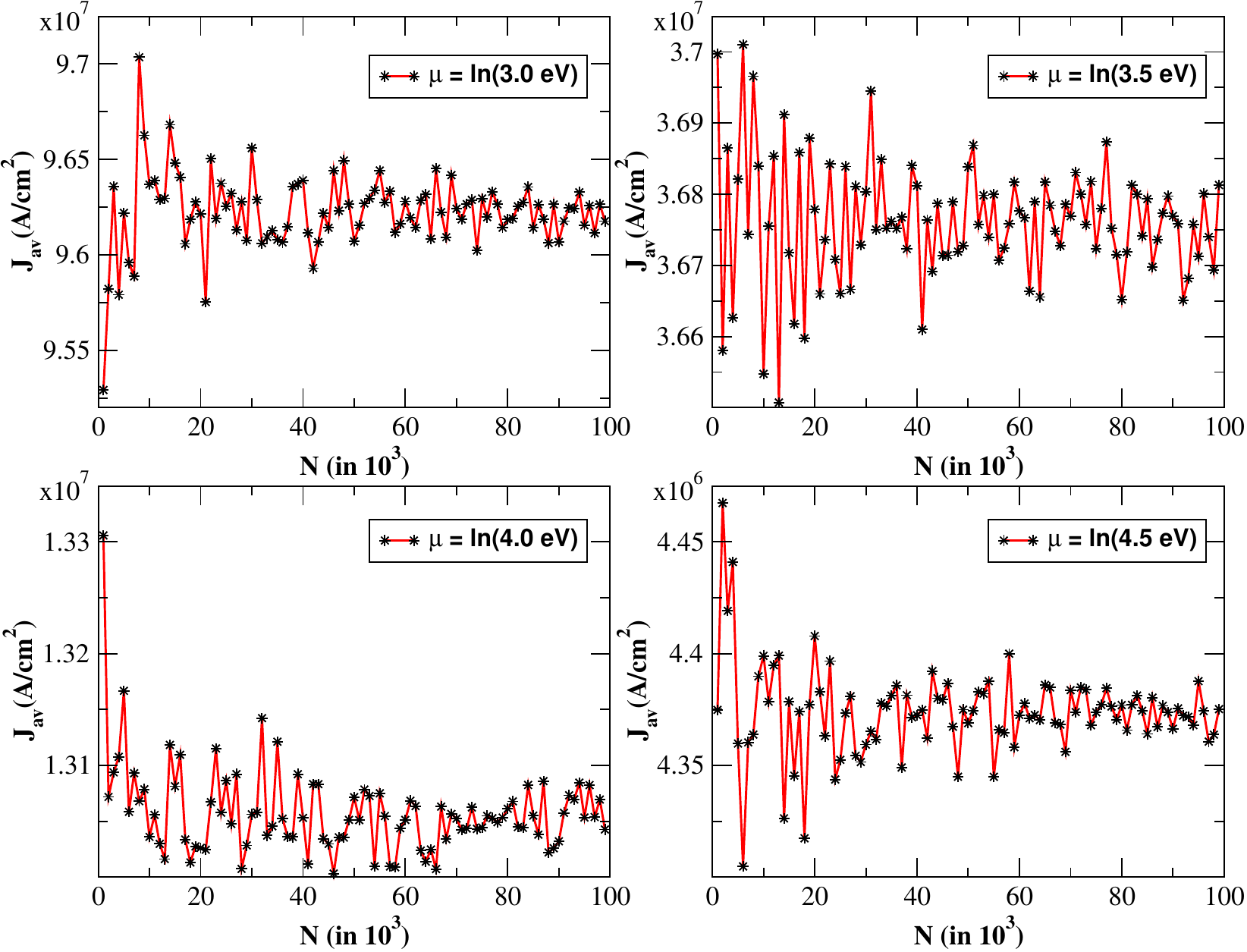}
\caption{(Color online)Average current ($J_{\textrm{av}}$) as a function of number of random variables(N) chosen from Log-normal distribution for 
$\Phi \gg E_{F}$.}
\label{Fig:JvsNEF0.05LogNorm}
\end{figure}

Finally we consider the case with $E_{F}=0.05$ eV and log-normal work function distribution. As in the other cases we first study the dependence of $J_{\textrm{av}}$ on the number of random 
variables $N$. As shown in Fig.~\ref{Fig:JvsNEF0.05LogNorm} $J_{\textrm{av}}$ shows strong fluctuations for $N < 10^{4}$ and settles to a mean value for $N > 6\times 10^{4}$. We choose $N=10^{5}$ 
for calculations. The magnitudes of currents are several orders of magnitude higher than the case of Gaussian distribution. However the currents are much smaller than the cases with 
$E_{F}=10.0$ eV and log-normal distribution.

\begin{figure}[htbp]
\centering
\includegraphics[clip,scale=0.32]{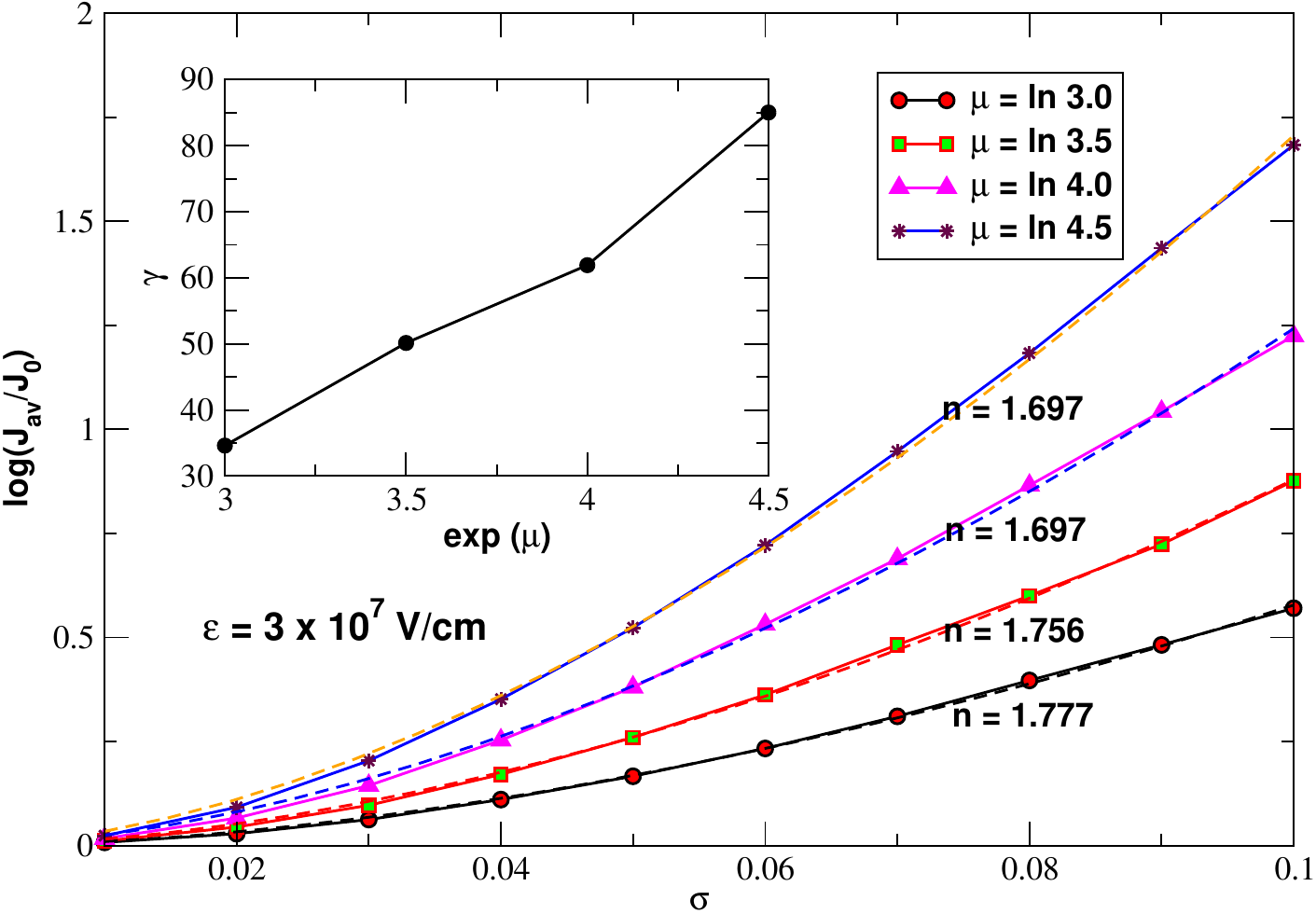}
\caption{(Color online) Logarithm of scaled average current, $J_{\textrm{av}}/J_{0}$, as a function of the width, $\sigma$, of the log-normal work function distribution for  $\Phi \gg E_{F}$. 
Solid lines with symbols are numerical data and the dashed lines represent fitting $g(\sigma)= \gamma\sigma^{n}$, where $n > 1$ is a non-universal exponent indicating compressed exponential 
behaviour of $J_{\textrm{av}}$. The value of $n$ for each case is indicated in the figure. Inset : variation of $\gamma$ with the median of the distribution $\exp(\mu)$. The applied field 
strength $\epsilon = 3\times 10^{7} \textrm{V/cm}$ and $E_{F}=0.05$ eV for all the cases.}
\label{Fig:JavVsSigmaMuEF0.05LogNorm}
\end{figure}

In Fig.~\ref{Fig:JavVsSigmaMuEF0.05LogNorm} we show $\log(J_{\textrm{av}}/J_{0})$ as a function of the width of the log-normal work function distribution for a given field strength 
$\mathcal{E} = 3\times 10^{7}$ V/cm for $\mu= \ln 3.0, \ln 3.5, \ln 4.0$ and $\ln 4.5$, respectively. This choice of $\mu$ corresponds to median value of the distribution $M = 3.0, 3.5, 4.0$ and 
4.5 eV, respectively. As in the earlier cases $J_{\textrm{av}}$ monotonically increases with $\sigma$. As in the earlier cases $\log(J_{\textrm{av}}/J_{0})$ can be fitted well with 
$g(\sigma)=\gamma\sigma^{n}$ with $n >1$. The fitting parameter $\gamma$ and the exponent $n$ are nearly identical with the case with $\Phi \ll E_{F}$. With increasing 
$M$, $\log(J_{\textrm{av}}/J_{0})$ increases and the exponent $n$ decreases. This is because with increasing $M$ ($\Phi$) the tuneling current decreases and thereby enhances the work function 
distribution effect and near exponental increase in tunneling current with $\sigma$. The fitting parameter $\gamma$ behaves linearly with $M = \exp(\mu)$

\begin{figure}[htbp]
\centering
\includegraphics[clip,scale=0.32]{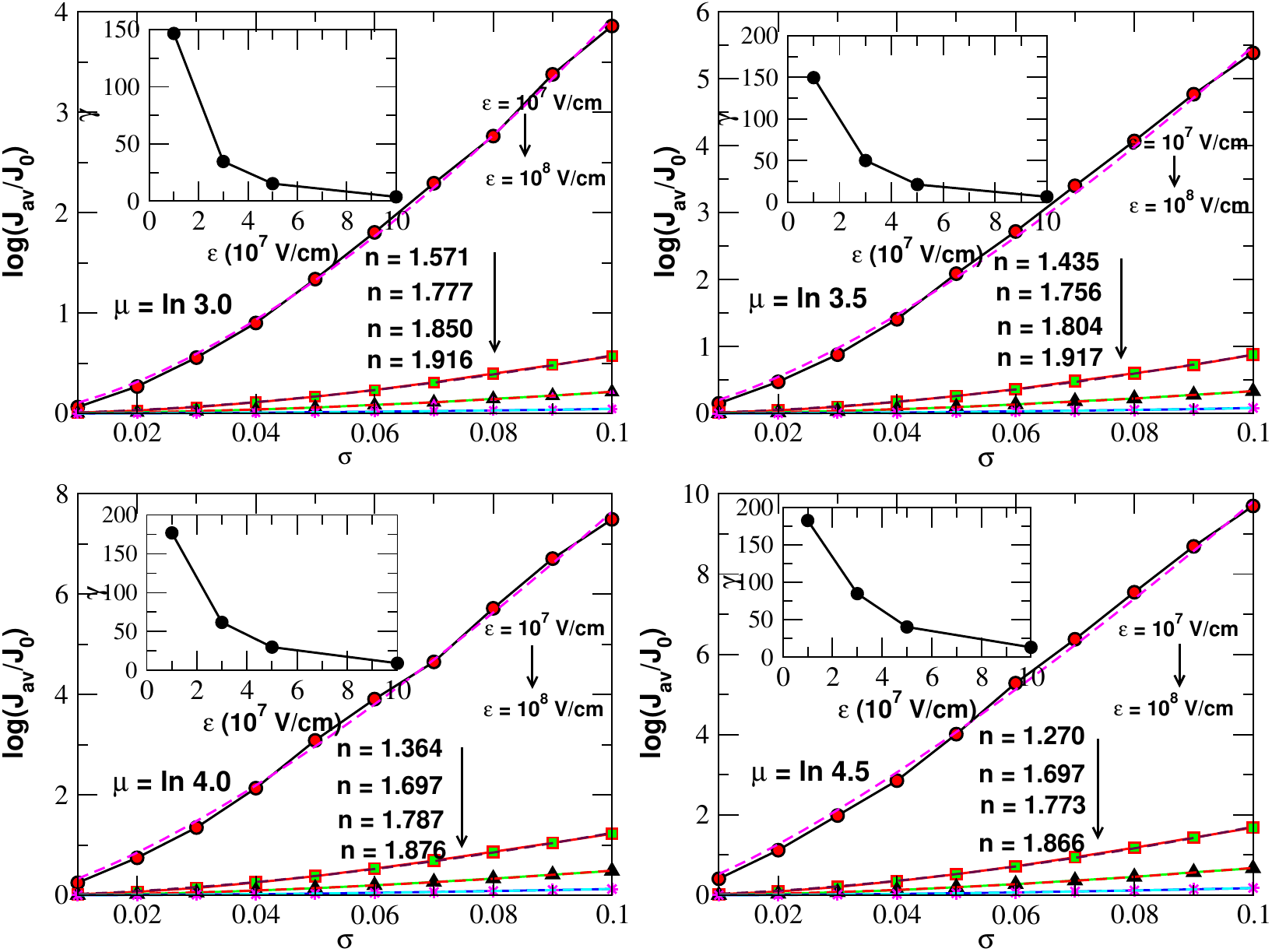}
\caption{(Color online) Panel plot : Logarithm of scaled average current $J_{\textrm{av}}/J_{0}$, as a function of the width, $\sigma$, of the log-normal work function distribution 
	for various applied field strength, $\mathcal{E}$. $\mathcal{E}$ varies from $10^{7} \textrm{V/cm}$ (top most curve) to $10^{8} \textrm{V/cm}$ (bottom most curve) in each plot. Solid lines 
with symbols are numerical data and the dashed lines represent the fitting function $g(\sigma)=\gamma\sigma^{n}$, where $n > 1$ is non-universal exponent. The value of $n$ for each 
curve is indicated in each plot. Inset of each plot shows variation of $\gamma$ with the median of distribution $\exp(\mu)$. Value of $\mu$ is explicitly shown in each plot and 
$E_{F}=0.05$ eV for each plot.}
\label{Fig:PanelPlotJavByJ0vsFEF0.05LogNorm}
\end{figure}

In Fig.~\ref{Fig:PanelPlotJavByJ0vsFEF0.05LogNorm} we show $\log(J_{\textrm{av}}/J_{0})$ as a function of the width, $\sigma$, of the log-normal work function distribution for various field 
strength, $\mathcal{E}$, as well as the median of the distribution $M=e^{\mu}$. For the lowest field strength $\mathcal{E}=10^{7}\textrm{V/cm}$ the tunneling current, $J_{\textrm{av}}$, averaged 
over the work function distribution is several orders of magnitude larger than $J_{0}$, the tunneling current calculated with $\Phi=M$. As stated earlier this is due to the fact that for 
$\mathcal{E}=10^{7}$ V/cm the tunneling current is very small and the skewness of the work function distribution effetively reduces the potential barrier significantly during the averaging process. 
With increasing field strength the tunneling current increases over several orders of magnitude and the effect of the work function distribution gets masked. With increasing $M$ the tunneling 
current decreases and the work function distribution effect enhances. For each case $\log(J_{\textrm{av}}/J_{0})$ can be fitted well with $g(\sigma)=\gamma\sigma^{n}$ with $n > 1$. 
The exponent, $n$, increases with the increasing field strength. For a given field strength, $\mathcal{E}$, $n$ decreases with increasing median value $M$ of the work function distribution. 
In particular, for $\mathcal{E}=10^{7}$ V/cm $n\sim 1$ giving rise to near exponential behaviour of the work function averaged tunneling current. The fitting parameter $\gamma$ sharply decreases 
with increasing field strength.

\section{Conclusions}
In conclusion, we have considered the role of work function distribution on the tunneling current in field emission effect. We have calculated the tunneling current, $J_{\textrm{av}}$, averaged 
over work function distribution. We have considered both Gaussian as well as a log-normal work function distribution. For each case we have studied dependence of $J_{\textrm{av}}$ on $N$, the 
number of random variables sampled over work function distribution. $J_{\textrm{av}}$ shows significant fluctuations for $N < 10^{4}$ but settles to an average value for $N > 4\times10^{4}$. 
We choose $N=10^{5}$ through out the calculation. We first calculate $J_{0}$, the tunneling current calculated for $\Phi=\Phi_{0}$(Gaussian) or $\mu = \ln M$ (log-norml), where $\Phi_{0}$ or $M$ 
(median of the distribution) corresponds to the \textit{bulk value} of the work function. For each case we have studied the logarithm of the scaled current distribution $\log(J_{\textrm{av}}/J_{0})$ 
as a function of the width, $\sigma$, of the work function distribution for various $\Phi_{0}$ (Gaussian) or $M$ (log-normal) and the applied field strength $\mathcal{E}$. We have also considered 
systems with high density characterized by $\Phi \ll E_{F}$ ($E_{F} = 10$ eV) as well as systems with low density characterized by $\Phi \gg E_{F}$ ($E_{F} = 0.05$ eV). Both for Gaussian and 
log-normal distribution $\log(J_{\textrm{av}}/J_{0})$ increases monotonically with the increasing width, $\sigma$, of the work function distribution for a given field strength $\mathcal{E}$. This is 
due to the fact that averaging over work function distribution leads to exploration of the regions of lower potential height and hence increased tunneling current. The dimensionless quantity
$\log(J_{\textrm{av}}/J_{0})$ can be fitted with $f(\sigma)=\alpha\sigma^{2}$ (Gaussian) or with $g(\sigma)=\gamma\sigma^{n}$ (log-normal). The fitting parameters $\alpha$ and $\gamma$ increases 
linearly with the bulk value of the work function i. e. $\Phi_{0}$ (Gaussian) or $M$ (log-normal). For log-normal distribution the non-universal exponent $n > 1$ shows compressed exponential behaviour. 
\subsection*{Acknowledgements}
We would like to thank Nei Lopes, Arghya Taraphder for many valuable discussions. One of us (N. P) would like to thank IIT, Kharagpur for local hospitality where part of the work was done. One of us 
(R. M) would like to thank Cetral Electronics Engineering Research Institute, Pilani for providing local hospitality and research support where part of the work was done.



\begin{thebibliography}{99}
\bibitem{FNpaper}{R. H. Fowler and L. Nordheim, Proc. Roy. Soc. London A, \textbf{119}, 173 (1928).}
\bibitem{MGpaper}{E. L. Murphy and R. H. Good, Phys. Rev., \textbf{102}, 1464 (1956).}	
\bibitem{ChoyJPCM2005}{T. C. Choy, A. H. Harker and A. M. Stoneham, J. Phys. Cond. Matt., \textbf{17}, 1505 (2005).}
\bibitem{CutlerSurfaceScience1993}{P. H. Cutler, J. He, J. Miller, N. M. Miskovsky, B. Weiss and T. E. Sullivan, Progress in Surface Science, \textbf{42},169185 (1993).}
\bibitem{ForbesUltramicro2001}{R. G. Forbes, K. L. Jensen, Ultramicroscopy, \textbf{89}, 1722 (2001).}
\bibitem{EdgcombePRB2005}{C. J. Edgcombbe, Phys. Rev. B, \textbf{72}, 045420 (2005).}
\bibitem{JensenAPL2006}{K. L. Jensen and M. Cahay, Appl. Phys. Lett., \textbf{88}, 154105 (2006).}	
\bibitem{FischerJVacSciTech2013}{A. Fischer, M. S. Mousa, and R. G. Forbes, J. Vac. Sci. Technol. B, \textbf{31}, 032201 (2013).}
\bibitem{ForbesJVacSciTech2013}{R. G. Forbes, A. Fischer, and M. S. Mousa, J. Vac. Sci. Technol. B,\textbf{31}, 02B103 (2013).}
\bibitem{KyritsakisProcRoySocA2015}{A. Kyritsakis and J. P. Xanthakis, Proc. R. Soc. A, \textbf{471}, 20140811 (2015)}
\bibitem{HolgatePRAppl2017}{J. T. Holgate and M. Coppins, Phys. Rev. Appl., \textbf{7}(4), 044019 (2017).}	
\bibitem{GamezJApplCrys2017}{XLiana Gamez, Maxwell Terban, Simon Billinge and Maria Martinez-Inesta, \textbf{50}, 741 (2017).}
\bibitem{LopesCondMat}{Nei Lopes and A. V. Andrade-Neto, arXiv:1408.3663v3}
\bibitem{LopesPhysLett2020}{Nei Lopes and A. V. Andrade-Neto, Phys. Lett. A, \textbf{384}, 126399 (2020).}
\bibitem{BiswasPhysPlasmas2017}{D. Biswas and R. Ramachandran, Phys. Plasmas \textbf{24}, 073107 (2017).}	
\bibitem{HaugBook}{A. Haug, \textit{Theoretical Solid State Physics}, Volume 1, (Pergamon Press, Oxford, 1975.)}	
\bibitem{ForbesAPL2006}{R. G. Forbes, Appl. Phys. Lett. \textbf{89}, 113122 (2006).}
\bibitem{ForbesJVacSciTech2010}{R. G. Forbes and J. H. B. Deane, J. Vac. Sci. Technol. B, \textbf{28}, C2A33 (2010).}
\bibitem{DolanPhysRev1953}{W. W. Dolan, Phys. Rev. \textbf{91}, 510 (1953).}
\bibitem{GradshteynBook}{I. S. Gradshteyn and I. M. Ryzhik, \textit{Tables of Integrals, Series and Products}. Academic, New York (1965).}	
\end{thebibliography}
\end{document}